\pdfoutput=1

\documentclass[11pt]{article}

\usepackage{emnlp2021}

\usepackage{times}
\usepackage{latexsym}

\usepackage[T1]{fontenc}

\usepackage[utf8]{inputenc}

\usepackage{microtype}
\usepackage{amsmath}
\usepackage{url}
\usepackage{xspace}
\usepackage{enumitem}
\usepackage{pifont}
\usepackage{amssymb}
\usepackage{pifont}
\usepackage{graphicx}
\usepackage{subfigure}
\usepackage[ruled,vlined]{algorithm2e}
\usepackage{tikz}
\usepackage{multicol}
\usepackage{multirow}
\usepackage{makecell}

\usepackage[colorinlistoftodos, textwidth=\marginparwidth,disable]{todonotes}

\newcommand{\cryptogru}{\textsc{CryptoGRU}\xspace}
\def\HiLi{\leavevmode\rlap{\hbox to \hsize{\color{yellow!20}\leaders\hrule height .8\baselineskip depth .5ex\hfill}}}
\newcommand{\cmark}{\text{\ding{51}}}
\newcommand{\xmark}{\text{\ding{55}}}
\newcommand*\circled[1]{\tikz[baseline=(char.base)]{
            \node[shape=circle,fill,inner sep=0.2pt] (char) {\textcolor{white}{#1}};}}
            
\newcommand{\CGone}{CG-\circled{1}\xspace}
\newcommand{\CGtwo}{CG-\circled{2}\xspace}

\title{\cryptogru: Low Latency Privacy-Preserving Text Model Inference}

\author{Bo Feng \and Qian Lou \and Lei Jiang \and Geoffrey Fox \\
  Intelligent Systems Engineering \\
  Luddy School of Informatics, Computing, and Engineering \\
  Indiana University Bloomington \\
  \texttt{\{fengbo,louqian,jiang60\}@iu.edu, gcf@indiana.edu} \\}

\begin{document}
\maketitle

\begin{abstract}
Homomorphic encryption (HE) and garbled circuit (GC) provide the protection for users' privacy. However, simply mixing the HE and GC in RNN models suffer from long inference latency due to slow activation functions. In this paper, we present a novel hybrid structure of HE and GC gated recurrent unit (GRU) network, \cryptogru, for low-latency secure inferences. \cryptogru replaces computationally expensive GC-based $tanh$ with fast GC-based $ReLU$, and then quantizes $sigmoid$ and $ReLU$ to smaller bit-length to accelerate activations in a GRU. We evaluate \cryptogru with multiple GRU models trained on 4 public datasets. Experimental results show \cryptogru achieves top-notch accuracy and improves the secure inference latency by up to $138\times$ over one of the state-of-the-art secure networks on the Penn Treebank dataset.
\end{abstract}

\section{Introduction}
\label{sec:intro}


Billions of text analysis requests are processed by powerful RNN models~\cite{BahdanauCB14neural,kannan2016smart} deployed on public clouds everyday. 
These text analysis requests contain private emails, personal text messages, and sensitive online reviews. 
For instance, Gmail smart reply generation needs to scan users' plaintext email messages anonymously~\cite{kannan2016smart}.


Prior work~\cite{juvekar2018gazelle} proposes a hybrid cryptographic scheme that uses homomorphic encryption (HE) to process linear layers and garbled circuits (GC) to compute activations in a convolutional neural network. Compared to convolutional neural networks (CNN), RNNs can achieve more competitive accuracy in text analysis tasks~\cite{BahdanauCB14neural,podschwadt2020classification,bakshi2020cryptornn}. Mixing HE and GC presents impressing results in secure classification tasks~\cite{barni2011privacy}. However, mixing HE and GC in RNN will suffer from a long inference latency due to the slow GC-based activations. In contrast to a CNN, a RNN~\cite{BahdanauCB14neural} requires more types of activations such as $tanh$ and $sigmoid$. The GC protocol~\cite{ohrimenko2016oblivious} has to use a huge garbled table to implement a $tanh$ or $sigmoid$ activation. Both garbling and evaluating such a large table add significant latency to RNN layers. Based on our experimental and theoretical analysis, the GC-based activations can occupy up to 91\% of the inference latency in a HE and GC hybrid secure GRU.

To reduce the GC-based activation latency, we propose a novel secure gated recurrent unit (GRU) network framework, \cryptogru, that achieves high security level and low inference latency simultaneously. We use SIMD HE kerel functions from~\citet{juvekar2018gazelle} to process linear operations in a GRU cell, while it adopts GC to compute activations. Our contributions are summarized as follows:
\begin{itemize}[noitemsep,topsep=0pt,leftmargin=*]
    \item We build a HE and GC hybrid privacy-preserving cryptosystem, \cryptogru, that uses HE operations to process multiplications and additions, and adopts GC to compute activations such as $ReLU$, $tanh$, and $sigmoid$.  
    \item We replace computationally expensive GC-based $tanh$ activations in a GRU cell with fast $ReLU$ activations without sacrificing the inference accuracy. We quantize GC-based $sigmoid$ and $ReLU$ activations with smaller bitwidths to further accelerate activations in a GRU.
    \item We implement all proposed techniques of \cryptogru and compared \cryptogru against state-of-the-art secure networks.
\end{itemize}

\section{Background and Related Work}

\paragraph{Text analysis using GRU.}
GRU and long short-term memory (LSTM) are two types of RNNs that can capture long term dependencies~\cite{ChungGCB14Empirical}, which are important text classification and text generation~\cite{BahdanauCB14neural}. A single LSTM cell has totally $4\times (n^2 + nm + n)$ parameters, while a single GRU cell has only $3\times (n^2+nm+n)$ parameters, where $m$ means the dimension of the input and $n$ for the dimension of the hidden state. 
Prior studies~\cite{ChungGCB14Empirical,BahdanauCB14neural} show GRU can the same level of inference accuracy as LSTM. 



\paragraph{Threat model and cryptographic primitives.}
We consider semi-honest corruptions~\cite{juvekar2018gazelle,lou2019she,chou2020privacy} in our threat model, where a server $S$ is hosting a model and many clients $C$ are sending inputs for inference using $S$' model. The client and the server adhere the protocol, but attempt to infer information about the other party's input. 
Our protocol hides model weights, biases, and activations of a network model, which are likely to be proprietary.

HE~\cite{gentry2009fully} is a cryptosystem that supports computation on ciphertext without decryption. 
GC enables two parties (Sender and Receiver) to jointly compute a function over their private data without revealing data beyond output from each other. A GC function is represented by a Boolean circuit with 2-input gates (e.g., XOR, AND, etc.). The Sender garbles the Boolean circuit and generates the garbled table. The Receiver receives the garbled table from an Oblivious Transfer~\cite{juvekar2018gazelle} and then evaluates the table. 
The total GC communication overhead is proportional to the number of non-XOR gates in the garbling function~\cite{rouhani2018deepsecure,riazi2018chameleon}.
For instance, a $12$-bit $ReLU$ requires only $30$ non-XOR gates, while a $12$-bit $tanh$ needs $>2K$ non-XOR gates. 
\paragraph{Comparison with prior privacy-preserving inference.}
Prior studies create GC-only~\cite{ohrimenko2016oblivious}, HE-only~\cite{chou2020privacy,badawi2019privft} and HE+GC hybrid~\cite{juvekar2018gazelle} privacy-preserving neural networks for secure inferences. GC-only secure networks have to pay huge communication overhead and long inference latency, whereas the HE-only networks cannot accurately implement nonlinear activations by only homomorphic multiplications and additions. So, secure networks~\cite{juvekar2018gazelle} implement linear layers with HE operations and nonlinear activations with GC operations.
We compare \cryptogru against prior related works in Table~\ref{t:t_tech_com}. PrivFT~\cite{badawi2019privft}, and FHE-Infer~\cite{chou2020privacy} are two HE-only secure neural networks.
While Gazelle~\cite{juvekar2018gazelle} is one of the first HE and GC hybrid convolutional neural networks, it does not support RNN cells. Although SHE~\cite{lou2019she} uses an emerging HE protocol (TFHE), many TFHE-based activations greatly prolong its inference latency in text analysis tasks. Several prior works HE-RNN~\cite{bakshi2020cryptornn,podschwadt2020classification} use HE to implement linear operations, and return the intermediate encrypted results to the client without non-linear operations. 

\begin{table}
\centering
\footnotesize
\setlength{\tabcolsep}{3pt}
\newcommand{\thiscellwidth}{12mm}
\begin{tabular}{|l|p{\thiscellwidth}|p{\thiscellwidth}|p{\thiscellwidth}|p{\thiscellwidth}|}
\hline
                          & Text tasks    & Accurate & Efficient & No decrypt \\ \hline
PrivFT                &  $\xmark$  &  $\xmark$  &  $\xmark$ &  $\cmark$ \\ \hline
\makecell[l]{FHE-Infer}                    &  $\xmark$  &  $\xmark$  &  $\xmark$ &  $\cmark$ \\ \hline
Gazelle                   &  $\xmark$  &  $\xmark$  &  $\xmark$ &  $\cmark$ \\ \hline
SHE                      &  $\cmark$  &  $\xmark$  &  $\xmark$ &  $\cmark$ \\ \hline
\makecell[l]{HE-RNN}         &  $\cmark$  &  $\cmark$  &  $\xmark$ &  $\xmark$ \\ \hline
\makecell[l]{\cryptogru} &  $\cmark$  &  $\cmark$  &  $\cmark$ &  $\cmark$ \\ \hline
\end{tabular}
\caption{The comparison of secure models. $\cmark$ means the scheme performs good under such condition or is friendly to the description and $\xmark$ means the opposite. }
\label{t:t_tech_com}
\end{table}

\paragraph{Latency Bottleneck and Motivation.}
In a typical GRU cell, there are nine stages of linear operations and two non-linear operations. In out baseline implementation using Gazelle, the non-linear operations take up to 91.37\%. Therefore GC-based non-linear operations are the bottleneck in this structure. This is further discussed in Section~\ref{s:cryptogru}.


\section{\cryptogru}\label{s:cryptogru}


\begin{figure*}
\centering
\includegraphics[width=.880\linewidth]{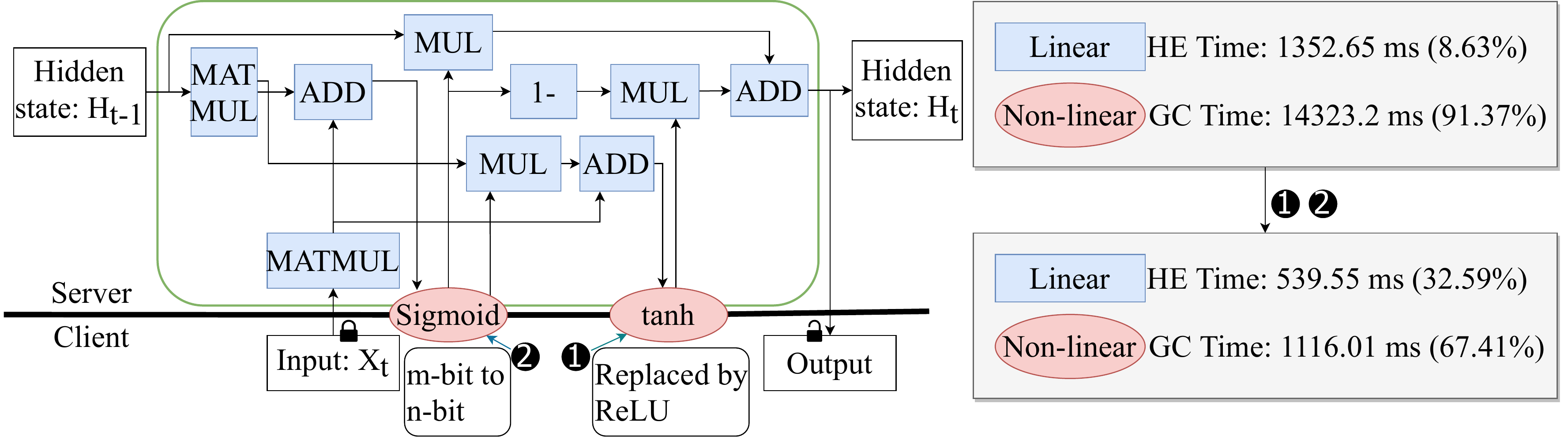}
\caption{A cell of \cryptogru with HE-based linear operations and GC-based non-linear operations.}
\label{fig:gru-wide}
\end{figure*}

\subsection{Constructing the base \cryptogru}
\label{s:con-base-gru}

Conventional neural network inference uses depth-bounded arithmetic circuits (LHE). However, the computation cost is large for the LHE scheme. \cryptogru adopts a simpler HE scheme, namely packed additive homomorphic encryption (PAHE) scheme and garbled circuits (GC).

\begin{algorithm}[ht!]
\footnotesize
\SetAlgoLined
\DontPrintSemicolon
\SetKwInput{KwData}{Input}
\KwData{an input ciphertext $[x_t]$}
\SetKwInput{KwData}{Output}
\KwData{a ciphertext hidden state $[h_t]$}
$[i_r], [i_i], [i_n] = MultPC([x], \Omega_{i}, b_{i})$ \tcp*{HE}
$[h_r], [h_i], [h_n] = MultPC([h_{t-1}], \Omega_{h}, b_{h})$ \tcp*{HE}
\HiLi{$[Gate_{reset}] = \mathbf{GCSig}(AddCC([i_r], [h_r]))$}\tcp*{GC}
\HiLi{$[Gate_{input}] = \mathbf{GCSig}(AddCC([i_i], [h_i]))$}\tcp*{GC}
\HiLi{$[Gate_{new}] = \mathbf{GCTanh}($} \tcp*{GC}
\HiLi{$~~AddCC([i_n], MultCC([Gate_{reset}], [h_n])))$}\tcp*{HE}
$[h_t] = AddCC([Gate_{new}],$ \tcp*{HE}
$~~MultCC([Gate_{input}], $ \tcp*{HE}
$~~~~AddCC([h_{t-1}], -[Gate_{new}])))$ \tcp*{HE}
\Return $[h_t]$
\caption{\cryptogru cell}\label{a:grucell}
\end{algorithm}
Figure~\ref{fig:gru-wide} illustrates the details of an internal view of a full GRU cell, which consists of both linear and non-linear operations. The linear operations are in blue, as shown in Figure~\ref{fig:gru-wide}. In a GRU cell, linear operations include matrix vector multiplications (\textit{MATMUL}), element-wise add, minus, and multiplications (\textit{ADD},\textit{1-},\textit{MUL}). Element-wise minus function is implemented as adding negative elements.
In \cryptogru, we map the neural network layers to PAHE matrix-vector multiplication for these linear operations.
The activation function $sigmoid$ and $tanh$ are non-linear, which are shown in red in Figure~\ref{fig:gru-wide}. For non-linear operations, we apply garbled circuits. The process of updating hidden states inside a GRU cell is shown in Algorithm~\ref{a:grucell}. Here, the $MultPC$ is a matrix vector multiplication based on HE, where the matrix is plaintext and the vector is ciphertext. We use $[~]$ to denote a ciphertext. In this function, $[i_r]$ is the product of $[x]$ with the first third of $\Omega_{i}$, $[i_i]$ is the product of $[x]$ with the second third of $\Omega_{i}$, and $[i_n]$ is the product of $[x]$ with the last third of $\Omega_{i}$. This mechanism also applies to the product of $[h_{t-1}]$ with $\Omega_h$. Here, the $AddCC$ and $MultCC$ are element-wise addition and multiplication respectively. In addition, the $GCSig$ and $GCTanh$ are the GC-enabled $sigmoid$ function and $tanh$ function respectively.





\subsection[tanh activation]{\circled{1} Replacing $tanh$ with $ReLU$}
\label{s:replace-tanh}


In this paper, we aim to build a privacy-preserving GRU network for text analysis. However, if we use the HE and GC hybrid technique~\cite{juvekar2018gazelle} to implement a GRU network, the complex GC-based activations including $tanh$ and $sigmoid$ significantly prolong the inference latency.
In GRU RNN, the activation nonlinearity function is typically $tanh$ but can also be implemented with the rectified linear unit $ReLU$~\cite{ravanelli2018light,ChungGCB14Empirical}. More importantly, we investigate that a $ReLU$ activation is more GC-friendly than a $tanh$ activation. A $8$-bit $ReLU$ activation requires $\sim 4\times$ less latency than a $8$-bit $tanh$ activation since a $8$-bit $ReLU$ requires only $24$ non-XOR gates, but a $tanh$ needs $95$ non-XOR gates.

\begin{table}
\centering
\footnotesize
\setlength{\tabcolsep}{3pt}
\begin{tabular}{|l|l|l|l|l|l|l|}
\hline
          & \multicolumn{2}{c|}{Accuracy} & \multicolumn{2}{c|}{Latency} \\ \hline
Datasets  & $tanh$          & $ReLU$         & $tanh$          & $ReLU$             \\ \hline
IMDB      & 84.8\%        & 84.6\%       & 14860ms              & \textbf{3779ms}               \\ \hline
Yelp Reviews & 77.3\%              &    78.1\%          &      5383ms         &       \textbf{1852ms}       \\ \hline
\end{tabular}
\caption{The $tanh$ activation vs. $ReLU$ activation.}\label{t:tanh_relu}
\end{table}

We use some tests against two public datasets to demonstrate this motivation. One dataset is the IMDB, which consists of 50,000 movie reviews. The other dataset is the Yelp review. Both datasets are used in binary classification tasks. In a one-layer GRU network, we compare the accuracy of using $tanh$ with $ReLU$ on the two datasets and compare the latency for a single sample inference. The results are summarized in Table~\ref{t:tanh_relu}. From this table, the GRU model with $ReLU$ can gain almost the same accuracy as that uses $tanh$ but trade off its training time for significant shorter latency during the inference stage. We label this version as ``\CGone'' in all the following text.

\subsection[quantize]{\circled{2} Quantizing both $sigmoid$ and $ReLU$}
\label{s:quantize}

During the computation of a full GRU cell, we identify the latency bottleneck is at non-linear functions.
In Figure~\ref{fig:gru-wide}, we show the computation time for non-linear operations hold about 91.37\% for a typical case.
This is mainly due to the computational complexity for ciphertext is significantly proportional to the underlying bit-length~\cite{riazi2018chameleon}. As shown in Table~\ref{t:tanh_relu}, the latency of $ReLU$ is significant less than that of $tanh$ due to simpler computational complexity~\cite{rouhani2018deepsecure}.

\begin{figure}
    \centering
    \includegraphics[width=.8\linewidth]{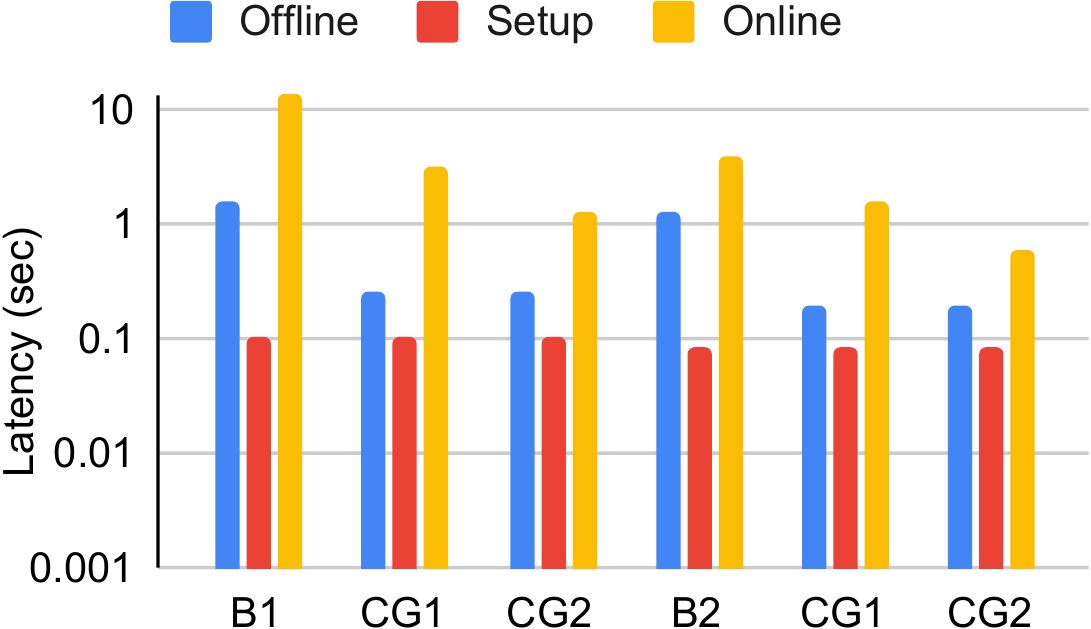}
    \vspace{-2mm}
    \caption{Latency comparison for two sets of experiments. ``B1'', the first ``CG1'', and the first ``CG2'' represents a model with the input size set at 10 and the hidden size set as 128. ``B2'', the second ``CG1'', and the second ``CG2'' represents a model with the input size set at 100 and the hidden size set as 64. ``CG1'' and ``CG2'' represents the \CGone~and \CGtwo~respectively. For both baseline cases, the online latency is significant. \CGone can reduce the online and offline latency and \CGtwo can further reduce the online latency while maintains the same level of the offline and setup.}
    \label{fig:perf-chart}
\end{figure}

Then, we quantize the all default bit-length from 20 to 8 in activations. 
First, the design of garbled circuits is proportional simpler after the quantization since the garbled circuits are sensitive to the bit length. Second, since these activation functions do not have any weight parameters, the overall accuracy of the neural networks with quantized activation functions can still hold.
We summarize the results of testing \cryptogru in Table~\ref{t:results-2} and we compare the latency shown in Figure~\ref{fig:perf-chart}. This benefit is further discussed in Section~\ref{s:exp}. We label this version as ``\CGtwo'' in all the following text.




\section{Experiments and Results}\label{s:exp}


\begin{table*}
\centering
\footnotesize
\setlength{\tabcolsep}{5pt}
\begin{tabular}{|l||r|r|r|r|}
\hline
Schemes         & \makecell{Input Size} & \makecell{Hidden Size} & \makecell{Total Latency} & \makecell{GC Msg Size}   \\\hline\hline
Baseline                  & 10 & 128 &  15675.85 ms & 213.4 MB \\\hline 
\CGone    & 10 & 128 & 3590.65 ms & 19.4 MB \\\hline
\CGtwo    & 10 & 128 & 1655.56 ms & 7.7 MB \\\hline
Baseline                  & 100 & 64  & 5382.91 ms & 106.7 MB \\\hline 
\CGone    & 100 & 64  & 1851.32 ms & 9.7 MB \\\hline
\CGtwo    & 100 & 64  & 913.32 ms & 3.9 MB \\\hline
\end{tabular}
\caption{The benchmark results of \cryptogru.}\label{t:results-2}
\end{table*}


\subsection{Cryptographic settings}
We develop the \cryptogru with the Gazelle SIMD Homomorphic operations in C++~\cite{juvekar2018gazelle}. Two main sets of cryptographic primitives are used for the \cryptogru inference. One set is for homomorphic encryption and the other set is for a garbled-circuit scheme. For the homomorphic encryption, we use Brakerski-Fan-Vercauteren (BFV) scheme~\cite{brakerski2012fully}. Yao's garbled circuits scheme is used for a two-party secure computation~\cite{yao1986generate}. We set the bit-length to 20 for plaintext and 60 for ciphertext in the BFV scheme as explained in Section~\ref{s:con-base-gru}.

\subsection[effectiveness]{Ablation studies from \circled{1} and \circled{2}}
We test our three versions of \cryptogru for the performance with respect to the latency. The results are shown in Table~\ref{t:results-2} and Figure~\ref{fig:perf-chart}. 
In a single GRU cell, there are 12 operations, 9 of which are linear and 3 are non-linear as discussed in Section~\ref{s:con-base-gru}. For each operation, we calibrate its setup latency, offline latency, online latency. In addition, the complexity of these operations are proportional to the size of input and configured hidden size. We use 30 time steps in the default settings. Here we illustrate two sets of input and hidden sizes. Given the input size is 10 and hidden size is 128, for the baseline case, the offline latency is 1651.2ms, the setup latency is 107.5ms, and the total latency is 1567.85ms. Compared to this case, \CGone can finish with a 258ms offline latency, 107.5ms setup latency, and 3225.15ms online latency, resulting a total of 3590.65ms latency. 

The offline latency and online latency are improved due to the simpler computational complexity of using $ReLU$. Benchmark results show that the $ReLU$ function can use 6.4 times less circuit gates for ciphertexts. This version is about 77\% faster than the baseline. By contrast, the \CGtwo has the same setup and offline latency as the \CGone, but the online latency is only about 1290.06ms, resulting that the total latency is 1655.56ms, which is about 54\% faster than the \CGone. The two techniques show the same effect when the input and hidden sizes are 100, and 64 respectively. The baseline version use a total of 5392.91ms, the \CGone use a total of 1851.32ms, and the \CGtwo use a total of 913.32ms. In this setting, the \CGone shows an improvement of about 66\% respect to the latency of the baseline and \CGtwo shows a further improvement of about 51\% compared to the \CGone.

Applying two techniques \circled{1} and \circled{2} can decrease the total latency and online message size for GC. Compared with related work, \cryptogru can achieve low latency in a secure inference system and maintain the same level of accuracy. There are some limitations due to the nature of homomorphic computing complexity. In addition, the recurrent computation would raise noise in homomorphic encryption. We mitigate the noise by bootstrapping the ciphertext~\cite{chillotti2020tfhe}.

\begin{table}
\centering
\footnotesize
\setlength{\tabcolsep}{3pt}
\begin{tabular}{|l||p{28mm}|r|r|}
\hline
Datasets                       & Neural Networks & \multicolumn{1}{l|}{Accuracy(\%)} & \multicolumn{1}{l|}{Latency} \\ \hline\hline
\multirow{2}{*}{\makecell[l]{Enron\\Emails}}  & PrivFT         & -                             & 7.95s$*$                       \\ \cline{2-4} 
                              & CryptoGRU       & 84.2                          & \textbf{2.03s}               \\ \hline
\multirow{2}{*}{\makecell[l]{Penn\\Treebank}} & SHE            & 89.8ppw                          & $\sim$576s                   \\ \cline{2-4} 
                              & CryptoGRU       & 79.4ppw                          & \textbf{4.14s}               \\ \hline
\multirow{3}{*}{IMDB}          & HE-RNN         & 86.47$*$                        & 70.6s$*$                       \\ \cline{2-4} 
                              & PrivFT        & 91.49$*$                        & 7.90s$*$                       \\ \cline{2-4} 
                              & CryptoGRU       & 84.6                          & \textbf{2.07s}               \\ \hline
\multirow{2}{*}{\makecell[l]{Yelp\\Review}}   & PrivFT         & 96.06$*$                        & 7.88s$*$                       \\ \cline{2-4} 
                              & CryptoGRU       & 91.3                          & \textbf{0.91s}               \\ \hline
\end{tabular}
\caption{Results from \cryptogru and related work.}\label{t:comp-results}
\end{table}

\subsection{Results}
\todo{Gazzle addressed x functions for CNN, however for RNN the case is different.}
We test the latency and accuracy of \cryptogru against public datasets and compare the performance with the state-of-art prior related works. We use Enron emails~\cite{klimt2004enron} and Penn Treebank datasets~\cite{le2015simple} that are common to machine learning tasks for text to evaluate the performance of the \CGtwo (referred as \cryptogru in this section) as well as the IMDB and Yelp datasets from Section~\ref{s:replace-tanh}. 
Experiments covered in Table~\ref{t:comp-results} are typical classification or regression tasks for text datasets.
We use the perplexity per word (PPW) as the target for the Penn Treebank dataset, which means the average log-probability per word. This is a common regression task for this dataset.
Enron Emails is a dataset collection consisting of 500,000 emails with subjects and body messages.For Enron email datasets, we classify emails as spam or ham. This is a binary classification task.
We perform the binary classification task for the IMDB dataset that labels the reviews either as positive or negative~\cite{maas-EtAl:2011:ACL-HLT2011}. 
For Yelp reviews dataset~\footnote{\url{https://www.ics.uci.edu/~vpsaini/}}, we also perform the binary classification task. Reviews with a star greater than and equal to 3 are regarded as positive.

We summarize the comparison results in Table~\ref{t:comp-results}. 
For the Penn Treebank dataset, our \cryptogru can infer a sample in 4.14s, which is about 138 times faster than the SHE~\cite{lou2019she}. For the IMDB datset, our \cryptogru can finish one sample inference within 2.07s on CPU, which is about 33 times faster than the HE-RNN~\cite{podschwadt2020classification}. The \cryptogru can infer a sample from Enron Emails in 2.03 and Yelp reviews in 0.91s.

\section{Conclusion}
Machine learning as a service attracts interest from many aspects in industry. 
Public cloud companies already launched prediction services. 
However, sending plaintext to model servers for inference raise attentions to user privacy issues. 
We propose \cryptogru, a secure inference building block for gated recurrent unit that emphasises on text-like or time series models. We elaborate all the improvements based on theoretical analysis and confirm the legitimacy for all optimization means. \cryptogru improves a GRU with homomorphic encryption, share secrets, and garbled circuits heterogeneously to achieve low latency as well as high accuracy.  

\subsection*{Code Availability}
\cryptogru code is available at: \url{https://github.com/bfeng/CryptoGRU}. The public repository also includes software dependencies like the `cryptoTools' and the `Gazelle' code. Used datasets from all experiments are downloadable from the internet as described in the text.

\section*{Acknowledgement}
The authors would like to thank the anonymous reviewers for their valuable comments and helpful suggestions. This work was partially supported by the National Science Foundation (NSF) through awards CCF-1908992, CCF-1909509, and CCF-2105972. This work is partially supported by the National Science Foundation (NSF) through awards CIF21 DIBBS 1443054, SciDatBench 2038007, CINES 1835598 and Global Pervasive Computational Epidemiology 1918626.

\bibliography{custom,refs}
\bibliographystyle{acl_natbib}

\end{document}